# Transition Metal Dichalcogenide 1T'-MoTe₂ Nanoscale Films as Spin Pumping Platforms

EMANUELE LONGO[1*], PINAKAPANI TUMMALA[2], M. BELLI[2], ALESSIO LAMPERTI[2], CHRISTIAN MARTELLA[2], ALESSANDRO MOLLE[2], MARCO FANCIULLI[3], ROBERTO MANTOVAN[2]

1. Institut de Ciència de Materials de Barcelona (ICMAB-CSIC), Campus UAB, Bellaterra, Catalonia 08193, Spain
2. CNR-IMM, Unit of Agrate Brianza, Via C. Olivetti 2, 20864 Agrate Brianza, Italy
3. Department of Material Science, University of Milano Bicocca, Via R. Cozzi 55, Milan 20125, Italy

*elongo@icmab.es  **roberto.mantovan@cnr.it

## Abstract

Transition Metal Dichalcogenides (TMDs) emerged as a promising class of materials to be used in spintronics, with the aim to promote efficient spin-charge conversion (SCC) in TMD/ferromagnet (FM)-based devices. The MoTe₂ semimetal with distorted orthorhombic crystal structure in the 1T' phase gathered particular attention because of its high spin-orbit coupling and reconfigurability as a type-II Weyl semimetal close to room temperature. Here, we report on the role of chemically grown 1T'-MoTe₂ thin films in inducing SCC in 1T'-MoTe₂/FM heterostructures as measured room temperature. Ferromagnetic resonance (FMR) and electrically detected spin pumping FMR measurements performed on 1T'-MoTe₂/Co/Au and 1T'-MoTe₂/Au/Co/Au heterostructures reveal spin mixing conductance value up to $\sim 1.6 \cdot 10^{20} \ m^{-2}$ and a spin hall angle of 1.7%. These finding position MoTe₂ thin films as competitive spin-charge converter option to compete with other functional materials (i.e., heavy metals, topological insulators), highlighting its potentiality for future applications in spintronic devices.

## Introduction

During the last decade Transition Metal Dichalcogenides (TMDs) positioned in the world of spintronic device as efficient spin-charge current converters and spin sink materials, characterized by competitive performances when compared to most-widely investigated heavy metals and topological materials.[1,2] Within the TMD family, tuning the spin orbit coupling (SOC) strength and band structure engineering are achievable by combining different chemical elements, thus offering an extended playground to optimize the spintronic properties in TMD/ferromagnet (FM) heterostructures. Among TMDs, the MoTe₂ semimetal with distorted orthorhombic crystal structure in the 1T' (monoclinic) phase gathered particular attention to be exploited in spintronic devices, because of its high SOC and temperature-driven structural transition to a type-II Weyl semimetal Td (orthorhombic) phase.[3,4]

In this work, we report on the spin charge conversion (SCC) performance of 1T'-MoTe₂ nanoscale films grown by chemical vapour deposition (CVD),[5] and structurally characterized by transmission electron microscopy (TEM), micro-Raman spectroscopy and atomic force microscopy (AFM) characterizations. Broadband ferromagnetic resonance (FMR) spectroscopy and spin pumping FMR (SP-FMR) are employed to study the role of 1T'-MoTe₂ thin films in inducing spin accumulation and the consequent SCC in MoTe₂/FM heterostructures, with 1T'-MoTe₂ thicknesses ranging from 6 to 20 nm.[6] Based on previous evidences in Te- and Se-based topological insulators,[6–10] the influence of Au spacers at the MoTe₂/Co interface is taken into account for the possible presence of topological surface states (TSS) driving the SCC mechanisms. Indeed, a transition from the trivial MoTe₂ 1T' phase to the topological 1T$_d$ one has been observed at temperature lower than 240K or for specific conditions of strain and/or chemical electron doping,[11] leading to the development of TSS in the form of Fermi arcs associated with the emergent Weyl semimetal phase.[3,12]

The evolution of the FMR signal linewidth as a function of resonant frequency reveals that the introduction of the 1T'-MoTe₂ layer significantly modifies the magnetization dynamics of an adjacent FM, enhancing the damping constant ($\alpha$) of the system as compared to the 1T'-MoTe₂-free reference. The $\alpha$ enhancement is interpreted within the framework of the spin Hall effect (SHE) and thus connected to the spin mixing conductance ($g_{mix}$), namely a quantity directly proportional to the generation of pure spin currents in the system. The extracted $g_{mix}$ value is comparable to those found in the most efficient spin-charge interconverter materials (i.e., $10^{19} - 10^{20}\ m^{-2}$), such as heavy metals and topological insulators,[6,13] and almost one order of magnitude larger than in MoTe₂- and WTe₂-based systems previously reported.[14,15] The efficiency of the SCC mechanism is also evaluated using the bulk-like spin Hall effect (SHE) model, from which a spin Hall angle (SHA) $\theta_{SHA} = 1.6\%$ is determined. The findings here presented underscore the potential of semi-metallic 1T'-MoTe₂ thin films for spintronic device applications at room temperature. Moreover, they demonstrate that the transfer of angular momentum at the 1T'-MoTe₂/FM interface can be further enhanced through the strategic incorporation of metallic interlayers, likely preserving non-trivial topology at the 1T'-MoTe₂ surface, as previously observed in other Te-based spin-charge converters.

**Results and Discussion**

1. **1T'-MoTe₂ chemical, structural and morphological characterization**

Figure 1a shows a high-resolution TEM image of a ~7 nm thick 1T'-MoTe₂ thin film grown by CVD on a p−SiO₂ substrate. The deposited material exhibits a highly crystalline structure with well-defined layered structure. TEM lamellae were acquired from nine different areas, revealing a uniform thickness of approximately 10 layers across the entire surface. The TEM analysis further confirms the excellent crystalline quality of the MoTe₂ films, down to a sharp interface with the SiO₂ substrate. Complementary energy-dispersive X-ray spectroscopy (EDX) analysis is presented in Figure S1 of the Supporting Information. The EDX spectra shown in Figure S1(b) were collected from the region highlighted by the yellow box in panel (a). The spectra exhibit the most prominent peaks corresponding to the L-subshell of Te at 3.79 keV and the L-subshell of Mo at approximately 2.29 keV. These findings confirm that MoTe₂ forms a uniform and well-defined overlayer on the SiO₂ surface, as further illustrated by the chemical maps of Mo and Te in Figures S1(c) and S1(d), respectively.

The morphology of the 1T'-MoTe₂ films was characterized through atomic force microscopy (AFM) measurements acquired on 4 x 5 µm² area. The AFM image shown in Figure 1 b clearly reveals the successful growth of a MoTe₂ film with a thickness of approximately 7 nm, in line with the TEM observation of an average of 10 layers across the sample area. The film appears compact and continuous over the scanned area, characterized by a granular morphology indicative of well-defined grains. The surface roughness is notably low, with a root-mean-squared (RMS) value below 2 nm, suggesting a high-quality film with minimal surface irregularities. This morphology and smoothness are consistent with uniform film deposition and excellent crystallinity at the nanoscale. Further details on the structural characterization and growth procedure of MoTe₂ used in this work are reported elsewhere.[5] Comparable crystalline quality is observed across all samples investigated in this study, confirming the reliability of the experimental protocol adopted to benchmark the SCC efficiency in these systems. To further investigated the structural properties of the grown material, vibrational phonon modes are studied through the acquisition of Raman spectra, as presented in Figure1c. The different (gray and blue) curves represent signals collected at different locations on the as grown sample, all revealing vibrational phonon modes at 108, 127, 178, 257, and 163 cm$^{-1}$. These peaks correspond to the $A_u$, $E_{1g}$, $A_g$, $A_{1g}$ of out-of-plane, and $B_g$ represent in-plane vibrational modes of the 1T'-MoTe₂ phase, respectively.[16] The distinct Raman peaks in Fig. 1c confirm the formation of a pure 1T' phase MoTe₂ across the entire 4 cm ×1 cm large area deposited using CVD, further supporting the high crystalline quality as probed by TEM (see Fig. 1a).

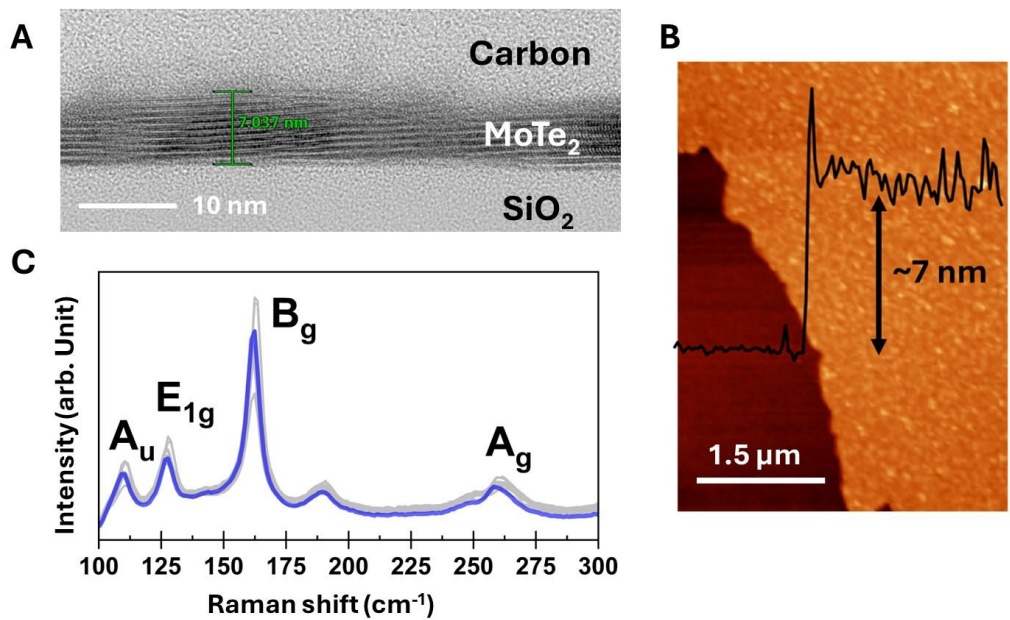

**Figure 1. A** Cross-sectional TEM image of MoTe2 films of thickness a 7nm (10 layers). **B** AFM image of the as-grown 1T'-MoTe₂ film and height profile. **C** Raman spectra of 1T'-MoTe₂ collected from 9 different locations of the sample area (gray and blue traces).

## 2. Influence of Au interlayer at the 1T'-MoTe$_2$/Co-based interface in inducing spin transparency

Figure 2 displays the comparison of the broadband FMR response between the 1T'-MoTe$_2$(10)/Co(5)/Au(5) and 1T'-MoTe$_2$(10)/Au(5)/Co(5)/Au(5) heterostructures, along with the corresponding Si/Co(5)/Au(5) and Si/Au(5)/Co(5)/Au(5) control samples. The numbers in brackets indicate the thickness of each layer in nm. In panel A the evolution of the Kittel curves acquired in the in-plane (IP) geometry is reported for each sample, where the resonant frequency ($f_{res}$) is plotted as a function of the resonant magnetic field ($H_{res}$).[6,17] For polycrystalline ferromagnetic thin films like ours (i.e., no IP anisotropy, see Methods), the Kittel equation has the form in Eq.1.

$$f_{res} = \frac{\gamma}{2\pi} \sqrt{H_{res}(H_{res} + 4\pi M_{eff})} \qquad (1)$$

were $\gamma = g \frac{e}{2m_e}$ the gyromagnetic ratio, with $g$ the g-factor, $e$ the charge of the electron, and $m_e$ the effective free electron mass. $M_{eff}$ represents the effective magnetization, accounting for the magnetic anisotropy of the system.[6,13,17] Since the Co layers are simultaneously deposited on all samples, we can reduce the number of free fitting parameters, keeping $\gamma$ constant at $1.98 \cdot 10^7 \frac{Hz}{Oe}$, which corresponds to the Landé factor g = 2.25.[18,19] The value of $M_{eff}$ extracted from each dataset shown in Figure 2a are reported in Table 1. All the samples shown a reduced $M_{eff}$ if compared with the Co bulk value (~1400 emu/cc), in accordance with surface shape anisotropy effect and/or magneto-structural disorder, as typically observed for thin films.[20,21]

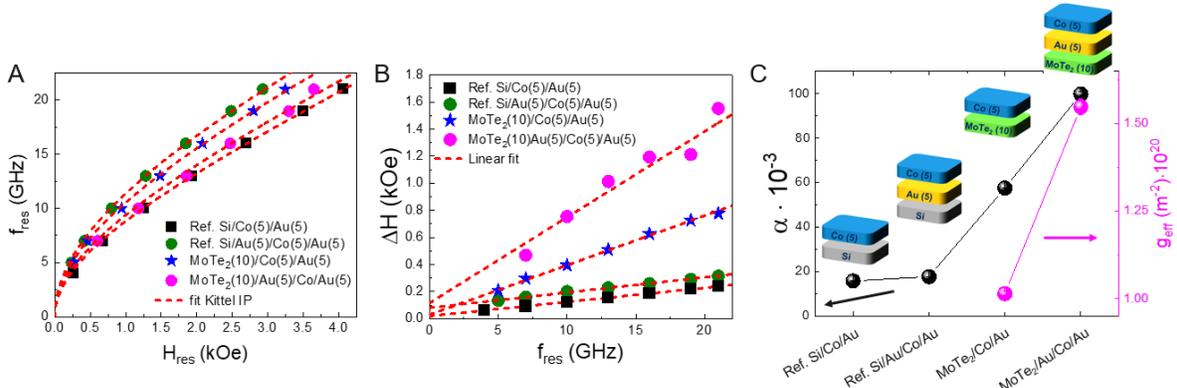

**Figure 2. A** Kittel dispersion for the samples reported in Table 1. The red dashed lines indicate the fit with Eq.1 **B** FMR signal linewidth as a function of the resonant frequency. The red dashed lines indicate the fit of the data with Eq.2, from which the damping parameter is extracted for each sample. **C** The black spheres indicate the damping constant (right y-axis) and the pink ones the spin mixing conductance values (right y-axis) for each sample of the dataset. The $g_{eff}^{\uparrow\downarrow}$ is extracted according to the expression reported in Eq.3.

**Table 1**: Parameters extracted from the datasets reported in Figure 2, for fixed $\gamma = 1.98 \cdot 10^7 \frac{Hz}{Oe}$.

| Stack | M$_{eff}$ (emu/cc) | $\alpha \cdot 10^{-3}$ | $\Delta H_0$ (Oe) | $g_{eff}$ ($\cdot 10^{20} \, m^{-2}$) |
|---|---|---|---|---|
| Si/Co(5)/Au(5) | 958 ± 11 | 15.8 ± 0.1 | 18 ± 1 | / |
| Si/Au(5)/Co(5)/Au(5) | 543 ± 5 | 17.7 ± 0.1 | 79 ± 2 | / |
| 1T'-MoTe$_2$(10)/Co(5)/Au(5) | 809 ± 9 | 57.5 ± 0.1 | 28 ± 7 | 1.01 ± 0.01 |
| 1T'-MoTe$_2$(10)/Au(5)/Co(5)/Au(5) | 626 ± 15 | 99.8 ± 1 | 115 ± 106 | 1.55 ± 0.02 |

In order to assess the capacity of 1T'-MoTe$_2$ to act as spin sink when in contact with FM, the linewidth ($\Delta H$) of the FMR signal is plotted as a function of $f_{res}$ and showed in Fig. 2B. Here, each dataset presents a linear dispersion, in agreement with the spin pumping model, and that can be fitted with the following equation

$$\Delta H = \Delta H_0 + \frac{4\pi\alpha}{\gamma} f_{res} \qquad [2]$$

where $\alpha$ indicates the damping constant (or Gilbert parameter) and $\Delta H_0$ the inhomogeneous broadening, a term accounting for magneto-structural disorder (i.e., oxidation) and/or texturization present in the FM layer. The control of the chemical-structural quality of the 1T'-MoTe$_2$(/Au)/Co interface is crucial for an efficient spin transport.[22] Thus, the values of $\Delta H_0$ should not exceed a critical threshold, usually well below 150-200 Oe as in the measured samples (see Table 1).[23] The relatively high $\Delta H_0$ value extracted for the 1T'-MoTe$_2$(10)/Au(5)/Co(5)/Au(5) dataset (pink circles) is ascribed to the scattered nature of two of the collected experimental point, not necessarily representing an indication of a lower magnetic quality of the Co layer in this sample (as we expect, being deposited simultaneously with the other samples).

In Figure 2C the black spheres indicate the $\alpha$ values as extracted by evaluating the slope of the linear curves (red dashed lines) in Fig. 2B (see Eq. 2). The correspondent numerical values are listed in Table 1.

In the framework of the spin pumping model, $\alpha$ is directly correlated to the accumulation of polarized spins in different areas of the Co layer, creating a flow of pure spin current. Then, such a flow of polarized spins is pumped into the neighboring 1T'-MoTe$_2$ layer, which acts as a spin sink. The spin transparency of the 1T'-MoTe$_2$/Co interface is evaluated through the real part of the spin mixing conductance ($g_{eff}^{\uparrow\downarrow}$), defined as

$$g_{eff}^{\uparrow\downarrow} = \frac{4\pi M_s t_{FM}}{g\mu_B}\left(\alpha_{1T'-MoTe_2} - \alpha_{Ref}\right) \qquad [3]$$

where $t_{FM}$ is the thickness of the FM layer and $\mu_B$ the Bohr magneton. $\alpha_{1T'-MoTe_2}$ and $\alpha_{1T'-MoTe_2-free}$ are the damping constant of the sample comprising the 1T'-MoTe$_2$ layer and their relative TMD-free reference, respectively. In panel C of Figure 2 the values of $g_{eff}^{\uparrow\downarrow}$ for the 1T'-MoTe$_2$(10)/Co(5)/Au(5) and 1T'-MoTe$_2$(10)/Au(5)/Co(5)/Au(5) sample are reported on the right y-axes, being $(1.01 \pm 0.01) \cdot 10^{20}\ m^{-2}$ and $(1.5 \pm 0.02) \cdot 10^{20}\ m^{-2}$, respectively. This analysis shows that the introduction of a 5 nm thick Au interlayer at the 1T'-MoTe$_2$(10)/Co(5) interface favors the overall spin transparency of the system, with $\alpha$ and $g_{eff}^{\uparrow\downarrow}$ enhanced of 66% and 50%, respectively, with respect the Au-free heterostructure. At first glance, this finding is counterintuitive, because the use of the Au spacer adds an additional interface to be crossed by the spin current generated in Co and pumped into the 1T'-MoTe$_2$ layer, which could cause additional and undesired scattering events (i.e., spin current backflow). Interestingly, similar improvement of both spin accumulation and consequent SCC effects, have been measured in Te-based topological insulator-based heterostructures upon introduction of Au interlayers.[6,7,9,10] The beneficial role of Au at the interface between Co and MoTe$_2$, can be understood based on thermodynamic considerations. Indeed, no energetically favorable reactions are foreseen between Mo and Au at room temperature and atmospheric pressure,[24,25] and similar arguments apply for Au vs. Te, with reported positive Gibbs free energy at the same thermodynamic conditions.[7] Simultaneously, Co is expected to efficiently "block" Au atoms' interdiffusion below 400 °C.[7,26] On the other hand, the direct contact of Co with the 1T'-MoTe$_2$ layer is suspected to be prone of a potentially very reactive interface, where Co-Te alloys could easily be formed as due to the considerably negative Gibbs free energy.[27–29]

We safely conclude that the 1T'-MoTe$_2$(10)/Au(5)/Co(5)/Au(5) heterostructure is characterized by stronger chemical stability and superior spin current generation efficiency than the 1T'-MoTe$_2$(10)/Co(5)/Au(5) counterpart, and this is the reason why we focus on the former system to seek for spin–charge conversion. In the following section, we also extend our study to 1T'-MoTe$_2$/FM heterostructures with varying 1T'-MoTe$_2$ thicknesses, in order to estimate the spin-diffusion length in MoTe$_2$.

### 3. Influence of Au interlayer at the 1T'-MoTe$_2$/Co-based interface in inducing spin accumulation

In Figure 3 the FMR thickness-dependent response of 1T'-MoTe$_2$(*t*)/Au(5)/Co(5)/Au(5) heterostructures is displayed, with *t* varying in the 0–22 nm range, where *t* = 0 indicates the 1T'-MoTe$_2$-free reference sample. As in the previous section, in panel A and B of Figure 3, the Kittel dispersion and the $\Delta H(f_{res})$ evolutions are reported, from which $M_{eff}$ and $\alpha$ are extracted and the numerical values collected in Table 2. To ensure a reliable comparison with the previous batch of samples, the data shown in Figure 3 for the Au-free Si/Au(5)/Co(5)/Au(5) reference and the 1T′-MoTe$_2$(10)/Au(5)/Co(5)/Au(5) samples are the same as those reported in Figure 2.

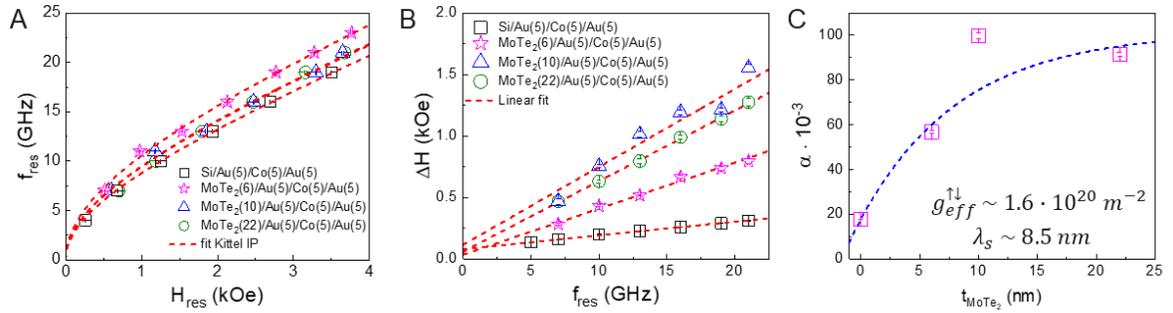

**Fig.3. A** Kittel dispersion for the samples with the Au interlayer and for different 1T'-MoTe$_2$ thicknesses, as reported in Table 2. Similar to Fig. 2, panel **B** shows the FMR linewidth vs resonant frequency evolution for samples in panel **A**. In panel **C**, the damping constant extracted from the linear fit in panel B is reported as a function of the 1T'-MoTe$_2$ layer thickness. From the latter dataset, $g_{eff}^{\uparrow\downarrow}$ is extracted using Eq. 3.

In Figure 3C, the $\alpha$ values as extracted from the fits in panel B are reported as a function of the thickness of the 1T'-MoTe$_2$ layer (pink empty squares), for each measured heterostructure. Here, $\alpha$ appears to saturate when 1T'-MoTe$_2$ exceed 10 nm, compatible with a predominant spin current accumulation generating due to bulk-like SHE, similarly to heavy metals (i.e., Pt).[30] The dataset in Figure 3C can be fitted writing Eq. 3 in terms of $\alpha$ and $t_{1T'\text{-MoTe}_2}$, having the form

$$\alpha = \frac{g\mu_B}{4\pi M_s t_{FM}} g_{eff}^{\uparrow\downarrow}\left(1 - e^{-\frac{t_{1T'-MoTe_2}}{\lambda_S}}\right) \qquad [4]$$

where $\lambda_S$ indicates the spin diffusion length in the 1T'-MoTe$_2$ layer.

**Table 2**: Parameters extracted from the datasets reported in Figure 3, for fixed $\gamma = 1.98 \cdot 10^7 \frac{Hz}{Oe}$. The sample marked with ** is the same one listed in Table 1, which is reported here for the reader's convenience.

| Stack | M$_{eff}$ (emu/cc) | $\alpha \cdot 10^{-3}$ | $\Delta H_0$ (Oe) |
|---|---|---|---|
| Si/Au(5)/Co(5)/Au(5) | $543 \pm 5$ | $17.7 \pm 0.1$ | $79 \pm 2$ |
| 1T'-MoTe$_2$(6)/Au(5)/Co(5)/Au(5) | $789 \pm 27$ | $56.7 \pm 0.1$ | $57 \pm 37$ |
| 1T'-MoTe$_2$(10)/Au(5)/Co(5)/Au(5) ** | $626 \pm 15$ | $99.8 \pm 1$ | $115 \pm 106$ |
| 1T'-MoTe$_2$(22)/Au(5)/Co(5)/Au(5) | $632 \pm 20$ | $99.8 \pm 0.11$ | $60 \pm 13$ |

From the fit to the data based on Eq. (4) in Figure 3C (blue dashed line) we infer $g_{eff}^{\uparrow\downarrow} = (1.63 \pm 0.45) \cdot 10^{20}\ m^{-2}$ and $\lambda_S = 8.5\ nm$. The $g_{eff}^{\uparrow\downarrow}$ value as derived from the thickness dependence in Fig. 3C is consistent with that reported in Section 1 via magnetodynamic response from a 10 nm-thick 1T'-MoTe$_2$ sample. On the other hand, a quantification of $\lambda_S$ for 1T'-MoTe$_2$ nanoscale film is still under assessment in the literature, with reported values scattered in a relatively broad range. For instance, an estimation of $\lambda_S$ is reported by Safeer et al.[31], being in the 1-20 nm range, too wide to be useful for practical applications. Conversely, Zheng et al.[15] quantified $\lambda_S$ through FMR experiments in 1T'-MoTe$_2$/NiFe heterostructure, being $13.96 \pm 1.74$ nm, similar to the ~14 nm extracted for 1Td-WTe$_2$.[14] The latter value was determined using similar FMR spectroscopy measurements in a vertical configuration, thus providing a valid comparison with our results. Although the value of 8.5 nm obtained in our case is comparatively lower, a fluctuation in $\lambda_S$ can be ascribed to the different crystalline quality of the deposited 1T'-MoTe$_2$ material, likely characterized by different chemical/structural defects. To be notice that, in our case, $g_{eff}^{\uparrow\downarrow}$ is two times larger than the value reported in Ref.[15] for similar systems, indicating a larger interface spin transparency.

To quantify the SCC efficiency the next section is dedicated to electrically detected SP-FMR measurements, where the 1T'-MoTe$_2$(10)/Au(5)/Co(5)/Au(5) heterostructure was chosen as representative system.

### 4. Spin-charge conversion in 1T'-MoTe$_2$/Au/Co/Au heterostructures

In Figure 4, the SP-FMR measurements for the 1T'-MoTe$_2$(10)/Au(5)/Co(5)/Au(5) (green empty circles) and Si/Au(5)/Co(5)/Au(5) (black empty squares) heterostructures are reported for a fixed $f_{res} = 10$ GHz. Here, on the y-axis, we report the two-dimensional (2D) charge current density ($J_c^{2D}$) as calculated by the voltage signal acquired across the samples, which is then normalized by the oscillating magnetic field ($h_{RF}$) exciting the magnetization of the FM layer (see Materials and Methods). In the framework of the SP-FMR theory, $J_c^{2D}(H_{ext})$ can be fitted using the following linear combination of Lorentzian functions[32,33]

$$J_c^{2D} = \frac{V_{mix}}{R \cdot W} = \frac{1}{R \cdot W}\left(V_{Sym}\frac{\Delta H^2}{\Delta H^2 + (H_{ext} - H_{res})^2} + V_{Asym}\frac{\Delta H(H - H_{res})}{\Delta H^2 + (H_{ext} - H_{res})^2}\right) \quad [5]$$

with $R$ and $W$, being the four-point resistance and the width of the sample. For the sample with the 1T'-MoTe$_2$ layer $R = 25$ Ω and $W = 2.3$ mm, while for the 1T'-MoTe$_2$-free reference $R = 25$ Ω and $W = 2.3$ mm. The DC voltage generated at the edges of a sample due to SP is commonly referred to as the mixing voltage ($V_{mix}$), as it results from the convolution of different physical phenomena. In general, $V_{mix}$ can be decomposed into its symmetric ($V_{sym}$) and antisymmetric ($V_{asym}$) components, which are typically attributed to the SP and spin rectification effects (i.e. Co layers' anisotropic magnetoresistance and/or anomalous Hall effect). The red solid lines in Figure 4 represent the fit of the data as obtained through Eq. 5. For both samples, the $J_c^{2D}$) curves are characterized by a highly symmetric shape (i.e., $V_{sym}$), indicating that spurious rectification contribution are suppressed in our systems (i.e., $V_{asym}$), while SP is the main origin of the signal. In order to get rid of thermal effects, the effective charge current density originated by SP is calculated as $J_c^{2D,SP} = \frac{1}{R \cdot W} \frac{V_{Sym}(+H_{ext}) - V_{Asym}(-H_{ext})}{2}$. The extracted $J_c^{2D,SP}$ values extracted from the fit in Figure 4 (red solid lines) are $0.042 \pm 0.001 \frac{mA}{m \cdot Oe}$ and $0.014 \pm 0.002 \frac{mA}{m \cdot Oe}$ for the 1T'-MoTe$_2$(10)/Au(5)/Co(5)/Au(5) and Si/Au(5)/Co(5)/Au(5) samples, respectively. The introduction of the 1T'-MoTe$_2$ layer produces a threefold increase of the charge current generated in the system, as a direct consequence of the SCC. According to the apparent plateau observed in Fig. 3C for the $g_{eff}^{\uparrow\downarrow}(t_{1T'-MoTe_2})$ curve, the SHE is the main mechanism dominating the SCC in our system, thus the SCC efficiency can be quantified through Eq. 6.

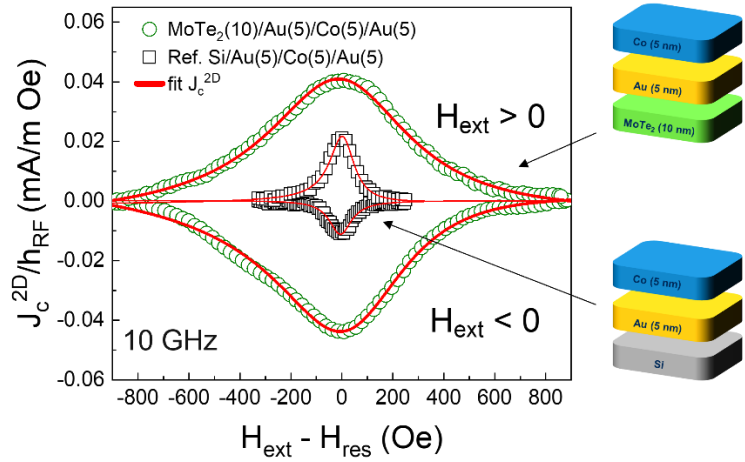

**Fig. 4**. SP-FMR signals as extracted from 1T'-MoTe$_2$(10)/Au(5)/Co(5)/Au(5) (green circles) and Si/Au(5)/Co(5)/Au(5) (black squares) heterostructures. The red solid lines represent the fit of the dataset with Eq. 5, from which the values of the generated charge current are extracted for each sample. The broader signal relative to the sample comprising 1T'-MoTe$_2$ layer (green circles) accounts for the enhanced damping constant in this sample with respect to the reference, while the increased intensity indicates the SCC induced by 1T'-MoTe$_2$.

$$\theta_{SHA} = \frac{J_c^{2D,SP}}{J_S^{3D} \lambda_s \tanh\left(\frac{t_{1T'-MoTe_2}}{2\lambda_s}\right)} \quad [6]$$

where $J_S^{3D}$ is the three dimensional (3D) pure spin current density generated in the system at resonance for h$_{RF}$ = 0.9 Oe (∝ RF-power), being $J_S^{3D} = 5.6 \cdot 10^5 \frac{A}{m^2}$ in our experimental conditions (see Supplemental Material).

By substituting all the quantities in Eq. 6, we extract $\theta_{SHA} = 0.016 = 1.6\%$, a value comparable with some values reported for the most efficient heavy metal spin charge converters as Pt, Ta and W.[34,35] Despite the $g_{eff}^{\uparrow\downarrow}$ and $\lambda_S$ values derived in the present work are comparable with those reported by Zheng *et al.*[15], the $\theta_{SHA}$ inferred from the data in Fig. 4 is comparatively lower, laying in their case into the 3 - 5% range. Interestingly, compared to our present work, in Ref.[15] the authors employed a different growth protocol to deposit their 1T'-MoTe$_2$ thin films. This could likely reflect a modified microstructural quality, possibly affecting differently the voltage acquisition during SP-FMR experiments due to inter-grains non-coherent electronic scattering.

The measured spin diffusion length and SCC efficiency measured in our 1T'-MoTe$_2$ are compatible with the presence of non-trivial topological phase, that can be originated by the formation of nuclei of MoTe$_2$ in the Td phase under strain effect or chemical electron doping in a thin 1T' matrix, as previously observed.[11,36]

**Conclusion**

In this work, TEM and Raman spectroscopy confirmed the high crystalline quality and the presence of the 1T′ phase in our MoTe$_2$ films, deposited by chemical vapor deposition following the protocol described by Tummala *et al.*[5]. Upon deposition of the Au/Co/Au ferromagnetic stack on top of 1T'-MoTe$_2$, FMR and electrically detected SP-FMR measurements were employed to investigate spin-charge interconversion effects in 1T′-MoTe$_2$. The FMR results demonstrate that 1T′-MoTe$_2$ acts as an efficient spin sink, exhibiting enhanced performance compared to common heavy metals. Furthermore, SP-FMR measurements reveal the ability of 1T'-MoTe$_2$ to convert pure spin currents into charge currents, with an efficiency comparable with similar TMD and heavy metal based spintronic heterostructures. Discrepancies have been found between the SCC performance of our systems compared with some previous reports, which we related to differences in the 1T'-MoTe$_2$ crystalline quality. So far, only a few reports on the spin-charge conversion mechanisms in 1T′- MoTe$_2$ have been published, leaving room for further investigation of this transition metal dichalcogenide as a potential material for spintronic applications. In this context, the present work constitutes a further step toward clarifying the spin-charge interconversion mechanisms and assessing the efficiency of 1T′- MoTe$_2$-based heterostructures.

**Materials and Methods**

MoTe$_2$ crystals were synthesized via chemical vapor deposition (CVD) on SiO$_2$ (50 nm)/Si substrates using a planarTECH system with a 2-inch diameter, 146 cm long quartz tube. The CVD setup consisted of two independently controlled heating zones. In the upstream zone, 100 mg of tellurium powder (99.998%, Sigma-Aldrich) was loaded, while 24 cm downstream, 1 mg of MoO$_3$ powder (99.98%, Sigma-Aldrich) was placed in a quartz boat. The boats had a semi-cylindrical shape (radius = 1 cm, height = 7 cm). The substrate was positioned face-down directly above the MoO$_3$. After evacuating the system to $3 \times 10^{-4}$ mbar, it was purged with 1000 sccm of high-purity argon. Growth was performed using distinct thermal ramp profiles (discussed below), with maximum temperatures of 800 °C (upstream) and 850 °C (downstream). After growth, the system was cooled to room temperature under 1000 sccm Ar flow. Further details can be found in Refs.[5,37].

Raman spectroscopy was conducted in backscattering geometry using a Renishaw InVia system with a 514 nm laser (2.41 eV) and a 50× Leica objective (NA = 0.75). The laser power was kept below 1 mW to avoid sample degradation.

The morphology of the samples was investigated in tapping mode using commercial AFM (Bruker Dimension Edge). Topographies were acquired in tapping mode using ultra-sharp silicon tips (TESPA-V2 Bruker radius of curvature 7 nm nominal frequency 320 kHz) Statistical parameters of the surface morphology, such RMS roughness, were derived by means of freely available software Transition electron microscopy was employed to assess the crystalline nature of the deposited $MoTe_2$ films in the cross-sectional configuration using a lamella of the sample. The lamellae preparation samples were prepared by means of Focused Ion Beam (FIB). The lamellae preparation was performed using a Thermo-Fischer Helios G4 FIB. In all the cases, particular care was taken to limit heating and ballistic effects of ion irradiation on $MoTe_2$ film during the final ion milling steps. After preparation, the lamellae were investigated by means of Scanning Transmission Electron Microscopy (STEM) techniques. The images were performed with a Thermo-Fischer Themis Z G3 aberration-corrected transmission electron microscope equipped with an electron gun monochromator operating at 200 kV acceleration voltage. Additionally, the chemical investigation of the sample was derived by means of STEM EDX technique operating at low beam current (0.5 nA).

The polycrystalline Co(5 nm)/Au(5 nm) bilayers and Au(5)/Co(5 nm)/Au(5 nm) trilayers were deposited on top of the $MoTe_2$ thin films via e-beam evaporation at RT. The polycrystalline nature of these stacks was identified through IP FMR measurements showing isotropic response, as reported in the supplemental material of Ref. [6]. The FMR experiments were conducted using a home-made setup, where the sample is positioned between the polar extensions of a Bruker ER-200 electromagnet with its surface parallel to the quasi-static magnetic field ($H_{ext}$) in the "flip-chip" configuration. For the FMR signal acquisition, the sample is fixed to a coplanar waveguide connected to a broadband (1 - 40 GHz) RF-source and the derivative of the absorption power downstream of the electrical transmission line is measured as a function of the $H_{ext}$ magnitude through a lock-in amplifier. The same experimental setup is equipped with a nanovoltmeter instrument to perform SP-FMR characterization. For further details on the FMR technique please refer to Refs.[6,9,13,38].

**Bibliography**
1. Galceran, R. *et al.* Control of spin-charge conversion in van der Waals heterostructures. *APL Materials* vol. 9 100901 Preprint at https://doi.org/10.1063/5.0054865 (2021).
2. Zibouche, N., Kuc, A., Musfeldt, J. & Heine, T. Transition-metal dichalcogenides for spintronic applications. *Ann Phys* **526**, 395–401 (2014).
3. Berger, A. N. *et al.* Temperature-driven topological transition in 1T'-MoTe2. *npj Quantum Materials 2018 3:1* **3**, 1–8 (2018).
4. Cheong, H., Cheon, Y., Lim, S. Y. & Kim, K. Structural Phase Transition and Interlayer Coupling in Few-Layer 1T′ and Td MoTe2. *ACS Nano* **15**, 2962–2970 (2021).
5. Tummala, P. P. *et al.* Large Area Growth and Phase Selectivity of MoTe2 Nanosheets through Simulation-Guided CVD Tellurization. *Adv Mater Interfaces* **10**, (2023).
6. Longo, E. *et al.* Large Spin-to-Charge Conversion at Room Temperature in Extended Epitaxial Sb2Te3 Topological Insulator Chemically Grown on Silicon. *Adv Funct Mater* **2109361**, (2021).
7. Longo, E. *et al.* Influence of Metal Interlayers on Spin-Charge Conversion in Sb2Te3 Topological Insulator-Based Devices. *Nano Lett* **25**, 6894 (2025).

8. Valant, M. *et al.* Chemical Interactions at the Interface of Au on Bi2Se3 Topological Insulator. *Journal of Physical Chemistry C* https://doi.org/10.1021/acs.jpcc.4c04241 (2024) doi:10.1021/acs.jpcc.4c04241.
9. Longo, E. *et al.* Spin-Charge Conversion in Fe/Au/Sb 2 Te 3 Heterostructures as Probed By Spin Pumping Ferromagnetic Resonance. *Adv Mater Interfaces* **2101244**, 2101244 (2021).
10. Fettizio, M., Avci, C. O., Mantovan, R. & Longo, E. Highly Efficient Current-Induced Torques Originating from Topological Surface States in Sb 2 Te 3. *Adv Electron Mater* https://doi.org/10.1002/aelm.202500280 (2025) doi:10.1002/aelm.202500280.
11. Paul, S. *et al.* Tailoring the phase transition and electron-phonon coupling in 1T′-MoTe2 by charge doping: A Raman study. *Phys Rev B* **102**, (2020).
12. Deng, K. *et al.* Experimental observation of topological Fermi arcs in type-II Weyl semimetal MoTe2. *Nat Phys* **12**, 1105–1110 (2016).
13. Longo, E. *et al.* Giant Spin-Charge Conversion in Ultrathin Films of the MnPtSb Half-Heusler Compound. *Adv Funct Mater* 2407968 (2024) doi:10.1002/ADFM.202407968.
14. Zheng, M. *et al.* Observation of Thickness-Modulated Out-of-Plane Spin–Orbit Torque in Polycrystalline Few-Layer Td-WTe2 Film. *Nanomaterials* **15**, 762 (2025).
15. Zheng, M. *et al.* Low-temperature fabrication, magnetoresistance and spin pumping studies of polycrystalline few-layer 1T'-MoTe2 films. *J Alloys Compd* **1015**, (2025).
16. Sun, Y. *et al.* Phase, Conductivity, and Surface Coordination Environment in Two-Dimensional Electrochemistry. *ACS Appl Mater Interfaces* **11**, 25108–25114 (2019).
17. Farle, M. Ferromagnetic resonance of ultrathin metallic layers. *Reports on Progress in Physics* **61**, 755–826 (1998).
18. Tokaç, M. *et al.* Interfacial Structure Dependent Spin Mixing Conductance in Cobalt Thin Films. *Phys Rev Lett* **115**, (2015).
19. Beaujour, J. M. L., Chen, W., Kent, A. D. & Sun, J. Z. Ferromagnetic resonance study of polycrystalline cobalt ultrathin films. *J Appl Phys* **99**, (2006).
20. Chappert, C., Dang, K. Le, Beauvillain, P., Hurdequint, H. & Renard, D. Ferromagnetic resonance studies of very thin cobalt films on a gold substrate. *Phys Rev B* **34**, 3192–3197 (1986).
21. Gladczuk, L., Aleshkevych, P., Lasek, K. & Przyslupski, P. Magnetic anisotropy of Au/Co/Au/MgO heterostructure: Role of the gold at the Co/MgO interface. *J Appl Phys* **116**, 233909 (2014).
22. Kaneta-Takada, S. *et al.* Enhancement of the spin hall angle by interdiffusion of atoms in Co2FeAl0.5Si0.5 / n - Ge heterostructures. *Phys Rev Appl* **14**, (2020).
23. Beaujour, J.-M., Ravelosona, D., Tudosa, I., Fullerton, E. E. & Kent, A. D. Ferromagnetic resonance linewidth in ultrathin films with perpendicular magnetic anisotropy. *Phys Rev B* **80**,.
24. Barin, I. Thermochemical Data of Pure Substances. *Thermochemical Data of Pure Substances* https://doi.org/10.1002/9783527619825 (1995) doi:10.1002/9783527619825.
25. Barin, I., Knacke, O. & Kubaschewski, O. Thermochemical properties of inorganic substances. *Thermochemical properties of inorganic substances* https://doi.org/10.1007/978-3-662-02293-1 (1977) doi:10.1007/978-3-662-02293-1.
26. Madakson, P. Interdiffusion, hardness and resistivity of Cr/Cu/Co/Au thin films. *J Appl Phys* **70**, 1374–1379 (1991).
27. Ahmed, S. *et al.* Doping and defect engineering induced extremely high magnetization and large coercivity in Co doped MoTe2. *J Alloys Compd* **918**, 165750 (2022).
28. Thajitr, W., Busayaporn, W. & Sukkabot, W. Electronic and magnetic properties of MoTe2 monolayer doped with single and double transition metals: Spin density functional theory. in *Journal of Physics: Conference Series* vol. 2431 (Institute of Physics, 2023).
29. Kumar, R. *et al.* Co doped Bi2Se3 topological insulator: a combined theoretical and experimental study. *Mater Sci Semicond Process* **195**, 109572 (2025).
30. Challab, N., Roussigné, Y., Chérif, S. M., Gabor, M. & Belmeguenai, M. Magnetic Damping and Dzyaloshinskii–Moriya Interactions in Pt/Co2FeAl/MgO Systems Grown on Si and MgO Substrates. *Materials* **16**, (2023).
31. Safeer, C. K. *et al.* Large Multidirectional Spin-to-Charge Conversion in Low-Symmetry Semimetal MoTe2 at Room Temperature. *Nano Lett* **19**, 8758–8766 (2019).
32. Tserkovnyak, Y., Brataas, A. & Bauer, G. E. W. Spin pumping and magnetization dynamics in metallic multilayers. *Phys Rev B Condens Matter Mater Phys* **66**, 1–10 (2002).

33. Feng, Z. *et al.* Spin Hall angle quantification from spin pumping and microwave photoresistance. *Phys Rev B* **85**, 214423 (2012).
34. Isasa, M., Villamor, E., Hueso, L. E., Gradhand, M. & Casanova, F. Erratum: Temperature dependence of spin diffusion length and spin Hall angle in Au and Pt [Phys. Rev. B 91, 024402 (2015)]. *Phys Rev B* **92**, 019905 (2015).
35. Chen, Y.-T. *et al.* Theory of spin Hall magnetoresistance (SMR) and related phenomena. https://doi.org/10.1088/0953-8984/28/10/103004 (2016) doi:10.1088/0953-8984/28/10/103004.
36. Tsipas, P. *et al.* Direct Observation at Room Temperature of the Orthorhombic Weyl Semimetal Phase in Thin Epitaxial MoTe2. *Adv Funct Mater* **28**, 1802084 (2018).
37. Martella, C. *et al.* Tailoring the Phase in Nanoscale MoTe2 Grown by Barrier-Assisted Chemical Vapor Deposition. *Cryst Growth Des* **21**, 2970–2976 (2021).
38. Georgopoulou-Kotsaki, E. *et al.* Significant enhancement of ferromagnetism above room temperature in epitaxial 2D van der Waals ferromagnet Fe5−δGeTe2/Bi2Te3 heterostructures. *Nanoscale* **15**, 2223–2233 (2023).

## Acknowledgments

E.L. acknowledge the funding provided by the Spanish Ministry of Science and Innovation through Projects PID2023-152225NB-I00 and Severo Ochoa MATRANS42 (CEX2023-001263-S), supported by MICIU/AEI/10.13039/501100011033 and FEDER, EU. E.L. also acknowledge support from Projects TED2021-129857B-I00 and PDC2023-145824-I00, funded by MCIN/AEI/10.13039/501100011033 and the European Union NextGeneration EU/PRTR, as well as from project 2021 SGR 00445 funded by the Generalitat de Catalunya. R.M. acknowledges financial support from the European Union's Horizon2020 project SKYTOP (FETPROACT 2018-01, no. 824123). Paolo Targa, Andrea Serafini and Davide Codegoni at the Physics Lab of STMicroelectronic Agrate Headquarter are acknowledged for the TEM measurements.

## Authors Contributions

E. L. conducted the FMR and SP-FMR experiments, data reduction and interpretation. M.B and E.L. developed the FMR and SP-FMR experimental setup under the supervision of R. M. and M. F.. P. Tummala, A. Lamperti, C. Martella produced the samples and the structural investigations. A. Molle supervised the samples production. M. Fanciulli supervised the activity at University of Milano-Bicocca. E. Longo and R. Mantovan conceived the study. E. Longo wrote the manuscript with contributions from all the authors.

# Transition Metal Dichalcogenide 1T'-MoTe₂ Nanoscale Films as Spin Pumping Platforms

EMANUELE LONGO[1*], PINAKAPANI TUMMALA[2], M. BELLI[2], ALESSIO LAMPERTI[2], CHRISTIAN MARTELLA[2], ALESSANDRO MOLLE[2], MARCO FANCIULLI[3], ROBERTO MANTOVAN[2]

1. Institut de Ciència de Materials de Barcelona (ICMAB-CSIC), Campus UAB, Bellaterra, Catalonia 08193, Spain
2. CNR-IMM, Unit of Agrate Brianza, Via C. Olivetti 2, 20864 Agrate Brianza, Italy
3. Department of Material Science, University of Milano Bicocca, Via R. Cozzi 55, Milan 20125, Italy

*elongo@icmab.es  **roberto.mantovan@cnr.it

*Keywords: spintronics, spin accumulation, transition metal dichalcogenide, Weyl semimetal, ferromagnetic resonance*

*Supplemental Material*

1. **EDS measurement on 1T'-MoTe₂ single layer thin film**

Energy-dispersive X-ray spectroscopy (EDX) analysis was performed in scanning transmission electron microscopy (STEM) mode to confirm the elemental composition and spatial distribution of the deposited layer. Figure S1 provides complementary EDX data: panel (a) shows the selected region of interest, highlighted by the yellow box, from which the spectra in panel (b) were acquired. The spectra display the characteristic L-subshell peaks of tellurium (Te) at 3.79 keV and molybdenum (Mo) at approximately 2.29 keV, indicating the presence of both elements in the analyzed area. This evidence supports the formation of a uniform and well-defined MoTe₂ overlayer on the SiO₂ substrate. The chemical maps of Mo and Te, shown in panels (c) and (d), further corroborate this conclusion by revealing a homogeneous spatial distribution of the two elements across the examined region.

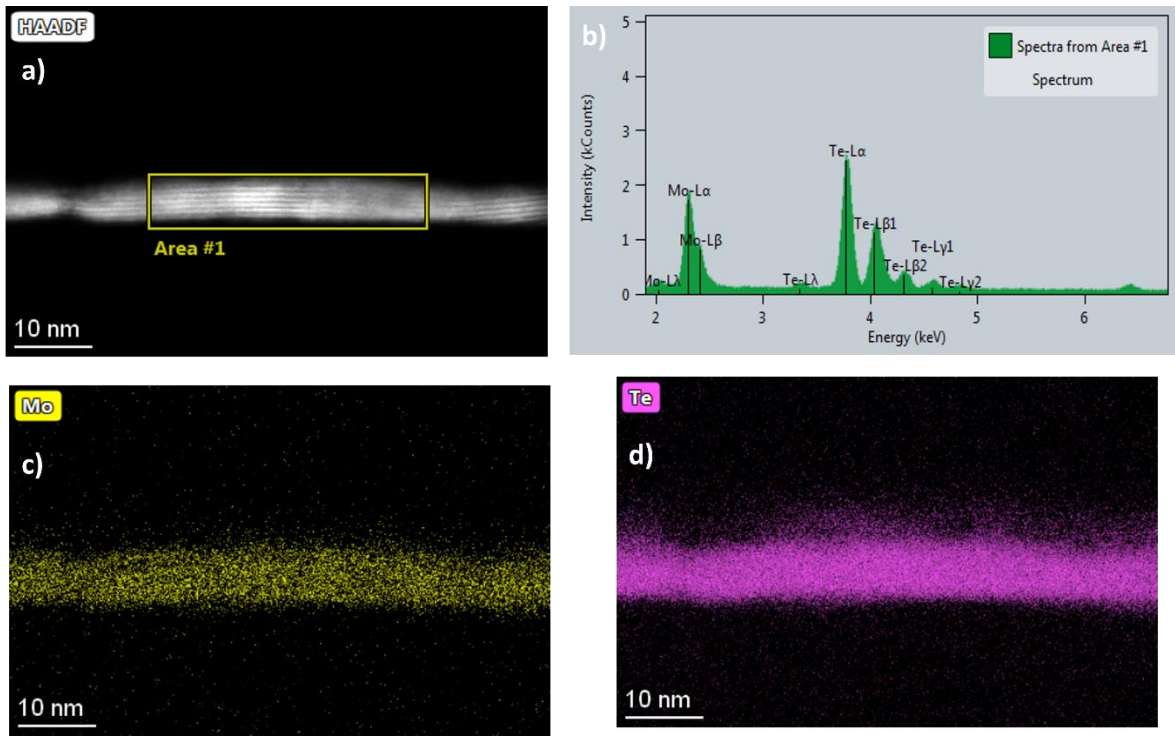

**Figure S1. TEM and EDX analysis of the MoTe₂ layers deposited on SiO2**

2. **Spin current density generated in 1T'-MoTe$_2$(10)/Au(5)/Co(5)/Au(5) heterostructure**

The electrically detected SP-FMR measurements are performed at RF-frequency of 10 GHz and an RF-power of 122 mW, corresponding to an $h_{RF}$ transverse oscillating magnetic field generated by the GCPW of 0.9 Oe. The generated 3D spin current $J_S^{3D}$ (in units of A/$m^2$) was calculated by using Equation S1.

$$J_S^{3D} = \frac{Re(g_{eff}^{\uparrow\downarrow})\gamma^2 h_{RF}^2 \hbar}{8\pi\alpha^2} \left(\frac{4\pi M_S\gamma + \sqrt{(4\pi M_S\gamma)^2 + 4\omega^2}}{(4\pi M_S\gamma)^2 + 4\omega^2}\right)\frac{2e}{\hbar} \qquad (6)$$

where $\hbar$ is the reduced Plank constant, $M_S$ the saturation magnetization, $\omega$ the frequency of the RF-signal, $e$ the charge of the electron. In our case we consider $M_S \sim M_{eff}$, thus to extract the $J_S^{3D}$ value reported in the main text we used $M_{eff} = 626 \pm 15$ emu/cc.